\documentstyle[12pt]{article}

\newcommand{\sect}[1]{\setcounter{equation}{0}\section{#1}}
\newcommand{\subsect}[1]{\subsection{#1}}

\def\be{\begin{equation}}
\def\ee{\end{equation}}
\def\bea{\begin{eqnarray}}
\def\eea{\end{eqnarray}}

\def\ww{{{\cal W}}}

\def\a{\alpha}

\def\om{\Omega}
\def\oom{{{\cal O}}}

\def\k{\omega}
\def\kk{\kappa}
\def\Ik{I_\k}

\def\diag{\,\mbox{diag}\,}
\def\ad{\mbox{ad}\,}

\def\>#1{{\bf #1}}                 
\def\<{\!<\!}

\def\1{\'{\i}}                           
\def\R{{\rm I\kern-.2em R}} 
\def\C{{\rm I\kern-.5em C}} 
\def\back{\!\!\!\!\!\!}

\begin{document}
\thispagestyle{empty}

\hfill \today
\ \vspace{2cm}

\begin{center}
{\LARGE{\bf{Casimir invariants for the complete}}}

{\LARGE{\bf{family of quasi-simple orthogonal }}}

{\LARGE{\bf{algebras}}}
\end{center}

\bigskip\bigskip

\begin{center}
Francisco J. Herranz${\dagger}$  and Mariano Santander${\ddagger}$
\end{center}

\begin{center}
\em 
{ { {}${\dagger}$ Departamento de F\1sica, Universidad de Burgos}
\\  E-09006, Burgos, Spain}

{ { {}${\ddagger}$ Departamento de F\1sica Te\'orica, Universidad de
Valladolid } \\   E-47011, Valladolid, Spain }
\end{center}
\rm

\bigskip\bigskip

\begin{abstract} 
A complete choice of generators of the center of the enveloping algebras
of real quasi-simple Lie algebras of orthogonal type, for arbitrary
dimension, is obtained in a unified setting. The results
simultaneously include the well known polynomial invariants of the
pseudo-orthogonal algebras
$so(p,q)$, as well as the Casimirs for many non-simple  algebras such as
the inhomogeneous $iso(p,q)$, the Newton--Hooke and Galilei type, etc.,
which are obtained by contraction(s) starting from the simple algebras
$so(p,q)$. The dimension of the center of the enveloping algebra of a
quasi-simple orthogonal algebra turns out to be the same as for the
simple $so(p,q)$ algebras from which they come by
contraction.  The structure of the higher order invariants is given in a
convenient ``pyramidal" manner, in terms of certain sets of
``Pauli--Lubanski" elements in the enveloping algebras.   As an example
showing this approach at work, the scheme is applied to recovering the
Casimirs for the (3+1) kinematical algebras. Some prospects on the
relevance of these results for the study of expansions are also given.
\end{abstract}

\newpage 


\sect{Introduction}

The role of Casimir (or polynomial) invariants of Lie algebras is rather
important in physics as well as in mathematics. They generate the center
of the universal enveloping algebra $U(g)$ of $g$. Physically, in any
theory with a symmetry algebra, they appear as related to conserved
quantities, as they commute with all generators. The work of Racah
\cite{Racah} solved the problem of obtaining the Casimir invariants
associated to a simple Lie algebra and, in particular,  Gel'fand
explicitly found a particular basis of the center of the enveloping
algebra of $so(N+1)$ \cite{Gelfand}; the case for the pseudo-orthogonal
Lie algebras $so(p,q)$ (in fact, for all simple classical algebras) has
been also dealt with in explicit form by Perelomov and Popov
\cite{PerelPop}. General results  concerning  both simple and non-simple
Lie algebras have been established; for instance, a formula giving the
number of primitive independent Casimir operators of any Lie algebra can
be found in
\cite{Beltrametti}, and a complete description of the theory of
polynomial and/or rational invariants appears in \cite{Abellanas}.
These results allow to deduce all invariants related to any particular
Lie algebra on a case-by-case basis, but do not give directly a general
perspective of the structure of the invariants associated to a complete
family of ``neighbour" algebras, as for instance provided
by a given Lie algebra and (some of) its contractions: the general problem
of relating the universal enveloping algebra of a given Lie algebra and one
of its possible contractions is not yet solved, though complete results
are available for special cases (see for instance, \cite{AbdelMalek}
where Casimirs for a large family of contractions of $sl(3, \C)$ are
given, or \cite{Montigny} where the problem of behaviour of bilinear invariants 
under contraction is studied within the graded contraction approach). 
This problem has definite interest in physics, where
contractions are related to some kind of ``approximation" and
understanding how invariants behave under contraction and under the
``inverse" expansion or deformation process are illuminating aspects of
the theory.

The aim of this paper is to obtain, within a such 
``simultaneous" approach, all the Casimir invariants for every Lie algebra
in a rather large family, the so-called quasi-simple or
Cayley--Klein (CK) algebras of orthogonal type \cite{Ros,Dubna96}. This
family includes the real simple pseudo-orthogonal Lie algebras $so(p,q)$
of the Cartan series $B_l$ and $D_l$ as well as many
non-simple Lie algebras which can be obtained by
contracting the former ones. The complete  family of
quasi-simple orthogonal algebras can be obtained
starting from the compact algebra
$so(N+1)$ in two different ways. One possibility is  to use a ``formal
transformation" which introduces numbers outside the real field
(either complex, double or dual (Study) numbers)
\cite{Grom}, and another makes use of the theory of graded
contractions \cite{MonPat,MooPat}, without leaving the real field.
Adopting this last point of view, it is shown that a particular
solution of the ${\bf Z}_2^{\otimes N}$ graded contractions of
$so(N+1)$  leads to the CK algebras as an $N$-parametric family of
real Lie algebras denoted $so_{\k_1,\dots,\k_N}(N+1)$
\cite{Grad}. The Lie algebra structure of the above family together
with a listing of its most interesting members are
briefly described in section 2.

When all $\k_a$ are different from zero, 
${so}_{\k_1,\dots,\k_N}(N+1)$ is a  simple algebra (isomorphic to 
$so(p,q)$ with $p+q=N+1$), whose rank is
$l=[\frac{N+1}{2}]$ (square brackets denoting here, as usual,  the integer
part). The dimension of the center of its universal enveloping algebra
equals the rank $l$ of the algebra, and it is generated by a set of 
homogeneous polynomials (Casimir operators) of orders $2, 4,\dots,
2[\frac{N}{2}]$, and an additional Casimir of order $l$ when $N+1$ is even.
We present in section 3 the explicit structure of the  Casimir invariants
corresponding to any algebra in the family $so_{\k_1,\dots,\k_N}(N+1)$.
These invariants  are deduced  starting from
the original approach of Gel'fand but where the necessary modifications are
introduced in order to get  expressions which cover simultaneously all
algebras in the family
$so_{\k_1,\dots,\k_N}(N+1)$, whether the constants $\k_a$ are
different from zero or not. This means that the behaviour of these
Casimirs upon any contraction $\k_a\to 0$ is built-in in the
formalism, and they do  not require any rescaling which should be made
when the contraction is performed in the In\"on\"u--Wigner sense. Every
Casimir we obtain is non-trivial for any contracted
algebra (whether or not the constants $\k_a$ are different from 
zero); furthermore, these  constitute a complete set of Casimirs
for CK algebras.

The main tool is provided by some elements in
the enveloping algebra, labeled by an even number $2s$  of indices, 
$W_{a_1a_2\dots a_sb_1b_2\dots b_s}$,   which are homogeneous of
order $s$ in the generators. In the case of the (3+1) Poincar\'e
algebra, the components of the Pauli--Lubanski vector (whose square
is the fourth-order Casimir) are in fact $W$-symbols  with four
indices.  In this way, the Casimir invariants are presented in a
pyramidal intrinsic form since each 
$W_{a_1a_2\dots a_sa_{s+1}b_1b_2\dots b_sb_{s+1}}$ can be written
in terms of $W$'s with two less indices
$W_{a_1a_2\dots a_sb_1b_2\dots b_s}$, and ultimately, in terms of
$W$-symbols  with two indices, which are simply the 
generators themselves. 

The problem of giving explicit expressions in terms of
generators for Casimirs in CK algebras has been also approached by
Gromov \cite{Gromov} by applying the above mentioned formal
transformation to the Casimir invariants of $so(N+1)$ obtained by
Gel'fand: those expressions should be equivalent to the ones we
obtain (as giving a possibly different basis for the center of the
enveloping algebra), but the explicit introduction of the
$W$'s makes the choice in this paper a lot simpler and easily tractable.
The general expressions for Casimirs written directly in terms of
generators, as in \cite{Gromov}, are overall more cumbersome to apply to
specific Lie algebras than the ones involving
$W$'s, specially  when $N$ increases. 

The results are illustrated in section 4 by writing the general
expressions  of the invariants associated to 
$so_{\k_1,\dots,\k_N}(N+1)$ for $N=2,3,4,5$; in particular, for
$N=4$ we  focus on  the (3+1) kinematical algebras \cite{BLL} thus
obtaining a global view of the limit transitions among their
corresponding Casimir operators. The way of getting the Casimirs
in the Minkowski space starting from those in   the DeSitter
space   by letting the universe ``radius" $R\to \infty$
is well known; this  familiar example appears in our scheme as a
rather particular case, yet it may facilitate grasping the scope of the
results we obtain, which includes a much larger family of algebras than
the well-known kinematical ones.

We make in the Conclusions some brief comments on the role of these
results for the study of expansions.


\sect{The family of quasi-simple orthogonal algebras}

Consider the real Lie algebra $so(N+1)$ whose $\frac 12 N(N+1)$
generators $\om_{ab}$ $(a,b=0,1,\dots, N, \  a<b)$ satisfy the following
non-vanishing Lie brackets: 
\be
[\om_{ab}, \om_{ac}] =  \om_{bc}  \qquad
[\om_{ab}, \om_{bc}] = -\om_{ac}  \qquad
[\om_{ac}, \om_{bc}] =  \om_{ab}  \qquad  a<b<c .
\label{aa}  
\ee
Through a ${\bf {Z}}_2^{\otimes N}$ graded contraction process a
family of contracted real Lie algebras can be deduced from
$so(N+1)$. The general solution has been given in 
\cite{SolGenGrad}; it includes from the simple Lie algebras
$so(p,q)$ to the abelian algebra of the same dimension. For reasons
which will become clear shortly, we restrict here to a particular
subfamily \cite{Grad}, whose members have been called quasi-simple
algebras \cite{Ros} because they are very ``near" to the simple
ones. They depend  on
$N$ real coefficients $\k_1,\dots,\k_N$ and the generic member of this
family will be denoted  $so_{\k_1,\dots,\k_N}(N+1)$. Their non-zero
commutators are given by 
\be
[\om_{ab}, \om_{ac}] =  \k_{ab}\om_{bc}  \quad\
[\om_{ab}, \om_{bc}] = -\om_{ac}  \quad\
[\om_{ac}, \om_{bc}] =  \k_{bc}\om_{ab}  \quad\  a<b<c  
\label{ab}  
\ee
without sum over repeated indices. Note that all Lie brackets
involving four different indices $a$, $b$, $c$, $d$ as $[\om_{ab},
\om_{cd}]$ are equal to zero.

The two-index coefficients $\k_{ab}$ are written in terms of the
$N$ basic $\k_{a}$ by means of:
\be
\k_{ab}=\k_{a+1}\k_{a+2}\cdots\k_b \qquad a,b=0,1,\dots,N \qquad a<b,
\label{ac}
\ee
therefore
\bea
&&\k_{a-1\, a}=\k_a \qquad a=1,\dots,N \label{ad}\\
&&\k_{ac}=\k_{ab}\k_{bc}  \qquad a<b<c.
\label{ae}
\eea
Each coefficient $\k_a$ can be reduced to $+1$, $-1$ or zero by a simple
rescaling of the initial generators; then  the family
$so_{\k_1,\dots,\k_N}(N+1)$ embraces
$3^N$ Lie algebras called CK algebras of orthogonal type or quasi-simple 
orthogonal algebras.   Some of them can be isomorphic; a useful isomorphism
is: 
\be
so_{\k_1,\k_2,\dots,\k_{N-1},\k_N}(N+1)\simeq
so_{\k_N,\k_{N-1},\dots,\k_2,\k_1}(N+1).
\label{af}
\ee
The algebra $so_{\k_1,\dots,\k_N}(N+1)$  has a (vector) representation by
$(N+1)\times (N+1)$ real matrices, given by
\be
\om_{ab}=-\k_{ab}e_{ab}+e_{ba}
\label{ccd}
\ee
where $e_{ab}$ is the matrix with a single  non-zero entry, 1, 
in the row $a$ and column $b$. In this realization, any element $X \in
so_{\k_1,\dots,\k_N}(N+1)$ satisfies the equation:
\be
X \, \Ik + \Ik\, {}^t X=0,
\label{cc}
\ee
where $\Ik$ is the diagonal matrix 
\be
\Ik=\diag (+,\k_{01},\k_{02},\dots,\k_{0N})=
\diag (+,\k_{1},\k_1\k_{2},\dots,\k_1\cdots\k_{N}) 
\label{ag}
\ee
and ${}^t X$ means the transpose matrix.
We state this property by saying that $X$ is an
$\Ik$-antisymmetric matrix; when all $\k_a=1$, this reduces to the
standard antisymmetry for the generators of $so(N+1)$.

The CK algebras $so_{\k_1,\dots,\k_N}(N+1)$ are the Lie algebras of
the motion groups of $N$-dimensional symmetrical homogeneous spaces
${\cal X}_0$:
\be
{\cal X}_0\equiv SO_{\k_1,\dots,\k_N}(N+1)/SO_{\k_2,\dots,\k_N}(N),\quad
\label{ah}
\ee 
where the subgroup $H_0\equiv SO_{\k_2,\dots,\k_N}(N)$  is generated by the
Lie subalgebra $h_0=\langle \om_{ab},\
a,b=1,\dots,N\rangle$. Each space ${\cal X}_0$ has
  constant curvature equal to $\k_1$ and its
principal metric can be reduced to the form $\diag
(+,\k_{2},\k_2\k_{3},\dots,\k_2\cdots\k_{N})$ at each point.

In the sequel we identify the most interesting Lie algebras
appearing within $so_{\k_1,\dots,\k_N}(N+1)$ according to the
cancellation of some $\k_a$ \cite{Tesis,extension}. In particular,   the
kinematical algebras \cite{BLL} associated to different models of
spacetime are CK algebras. In the list below, when we explicitly
say that some coefficient is equal to zero   it will be
understood that the remaining ones are not. It is remarkable that each case
$\k_a=0$ can be regarded as an
In\"on\"u--Wigner contraction limit, where some parameter 
$\varepsilon_a\to 0$ \cite{Grad,IW}.

\noindent
{\bf (1)} $\k_a\ne 0$ $\forall a$. They are the pseudo-orthogonal
algebras $so(p,q)$ with $p+q=N+1$ of the Cartan series $B_l$ or
$D_l$. The quadratic form invariant under the fundamental vector
representation is given by the matrix
$\Ik$ (\ref{ag}). 

\noindent
{\bf (2)} $\k_1=0$.   They are inhomogeneous pseudo-orthogonal
algebras $iso(p,q)$ with $p+q=N$ which have a semidirect sum
structure:
$$
so_{0,\k_2,\dots,\k_N}(N+1)
\equiv t_N\odot  so_{\k_2,\dots,\k_N}(N)\equiv iso(p,q) .  
$$
The ``signature" of the metric invariant under $so_{\k_2,\dots,\k_N}(N)$ is
$(+,\k_{12},\k_{13},\dots,\k_{1N})$.
The most interesting cases are the Euclidean algebra $iso(N)$ which
is recovered once for
$(\k_1,\k_2,\dots,\k_N)={(0,+,\dots,+)}$, 
and the Poincar\'e algebra  $iso(N-1,1)$ which appears several
times,  for instance, for    
$(\k_1,\k_2,\dots,\k_N)=\{(0,-,+,\dots,+)$, 
$(0,+,\dots,+,-)$, $(0,+,\dots,+,-,-,+,\dots,+)$,
$(0,-,-,+,\dots,+)$,  $(0,+,\dots,+,-,-)\}$. 
From the isomorphism (\ref{af}) is clear that the CK algebras with
$\k_N=0$ are also of this kind and similarly for the next types.
 
\noindent
{\bf (3)} $\k_1=\k_2=0$. They are twice
inhomogeneous pseudo-orthogonal algebras $iiso(p,q)$ with $p+q=N-1$:
$$
so_{0,0,\k_3,\dots,\k_N}(N+1)\equiv t_N\odot  \left( t_{N-1}\odot  
so_{\k_3,\dots,\k_N}(N-1)\right)\equiv iiso(p,q) .   
$$
The signature of  $so_{\k_3,\dots,\k_N}(N-1)$ is
$(+,\k_{23},\k_{24},\dots,\k_{2N})$. Hence the Galilean algebra 
$iiso(N-1)$ is associated to $(0,0,+\dots,+)$.  
 
\noindent
{\bf (4)} $\k_1=\k_N=0$. 
They are $ii'so(p,q)$ algebras with $p+q=N-1$:
$$
so_{0,\k_2,\dots,\k_{N-1},0}(N+1)\equiv t_N\odot({t'}_{\!N-1}\odot
    so_{\k_2,\dots,\k_{N-1}}(N-1))\equiv ii'so(p,q),
$$
where $so(p,q)$ acts on $t_N$ through the vector representation
while  it acts on ${t'}_{ N-1}$ through the contragredient of the
vector representation. Thus,   $ii'so(N-1)$ is the Carroll algebra
with coefficients $(0,+,\dots,+,0)$ \cite{BLL}.

\noindent
{\bf (5)} $\k_a=0$, $a\neq 1,N $. 
They are the $t_r(so(p,q)\oplus so(p',q'))$ algebras \cite{WB}. In
particular, for $\k_2=0$ we have $t_{2N-2} (so(p,q)\oplus
so(p',q'))$ with $p+q=N-1$ and $p'+q'=2$, which include  the
expanding and   oscillating Newton--Hooke algebras for $q=0$
\cite{BLL}.  
 
\noindent
{\bf (6)} 
When {\em} all coefficients $\k_a=0$ we find the flag space algebra 
$so_{0,\dots,0}(N+1)\equiv i\dots iso(1)$ \cite{Ros}.
 

\sect{The Casimir invariants}

We first recall the necessary definitions and tools
\cite{Beltrametti,Abellanas}. Then  we  summarize the approach and the
results of Gel'fand  for  $so(N+1)$ \cite{Gelfand}. Afterwards we
compute the Casimir invariants for the CK algebras
$so_{\k_1,\dots,\k_N}(N+1)$; when there is no risk of confusion we
shall denote this Lie algebra simply as $g$.

\subsect{Definition of polynomial invariants}

Let  $Ug$ the  {\it enveloping algebra} of $g$ generated by all
polynomials in $\om_{01},\dots,\om_{N-1N}$ and  $S$ the  {\it 
symmetric algebra} of $g$ isomorphic to 
${\R}[\a_{ab};\,a,b=0,1,\dots,N;\,a<b]$, this is, the ring of
polynomials in $\frac 12 {N(N+1)}$ commutative variables $\a_{ab}$.
A generic polynomial is denoted as
$p=p(\a_{01},\dots,\a_{N-1N})$. 

The {\it adjoint action}, $\ad \om_{ab}: g\longrightarrow  g$   
\be
\ad \om_{ab}(\om_{cd})=
[\om_{ab},\om_{cd}]  ,
\label{ba}
\ee
 is extended to an ``adjoint action" of $g$ on 
$Ug$ and also to another action on $S$:
\be
\ad \om_{ab}\ :\ u\in Ug \longrightarrow 
[\om_{ab},u]\equiv \om_{ab} u- u
\om_{ab}\in Ug,
\label{bb}
\ee
\be
\ad \om_{ab}\ :\ p=p(\a_{01},\dots,\a_{N-1N})\in S 
\longrightarrow
{\oom}_{ab}(p)\equiv { \sum_{
 c,d,m,n=0}^N} C_{ab,cd}^{mn}\a_{mn}
\frac{\partial p}{\partial \a_{cd}} \in S,
\label{bc}
\ee
where $C_{ab,cd}^{mn}$ are the structure constants of $g$. We are
using the pairs $ab$ as indices for the basis elements of $g$, so
the conditions $a<b$, $c<d$ and $m<n$ will be assumed without
saying. The structure constants themselves, read from
(\ref{ab}), are:
\be
C_{ab,ac}^{mn}=\delta_{mb}\delta_{nc}\k_{ab}, \quad
C_{ab,bc}^{mn} = - \delta_{ma}\delta_{nc},  \quad
C_{ac,bc}^{mn} =\delta_{ma}\delta_{nb}\k_{bc}, \quad a<b<c .
\label{bd}
\ee

The invariants in $Ug$ and $S$ under the adjoint action of $g$ are
the following subsets:
\bea
U^I  g&\equiv&\{u\in U g\ |\ [\om_{ab},u]=0,\ \forall
 \om_{ab}\in  g\}\subset
U g,\label{be}\\ 
S^I&\equiv&\{p\in S\ |\ {\oom}_{ab}(p)=0,\ \forall
\om_{ab}\in  g\}\subset S,
\label{bf}
\eea
and the elements of $U^I g$ are called 
{\em polynomial} or {\em Casimir invariants} of $g$.
According to the general results described in
\cite{Beltrametti,Abellanas} the two main steps to obtain the Casimir
invariants are:

\noindent
$\bullet$ To compute the subset   $S^I$ of $S$.
\smallskip

\noindent
$\bullet$ To apply to each element of $S^I$ the canonical
mapping  $\phi: S\to U g$ defined for any monomial  by symmetrization:
\be
\phi(\a_{a_1b_1}\dots\a_{a_rb_r})=\frac 1{r!}\sum_{\pi\in
\Pi_r}\om_{\pi(a_1b_1)}\cdots \om_{\pi(a_rb_r)},
\label{bg}
\ee
where $\Pi_r$ is the group of permutations on  $r$ items, and extended to
$S$ by linearity.
  
An upper bound for the number of
independent invariants for $g$ is provided by:

\noindent
{\it Proposition 1} \cite{Abellanas}. The maximal number of algebraically
independent Casimir invariants  $\tau$ associated to any algebra
$g$ is 
\be
\tau \le {\mbox{dim}}\, (g) -r(g),
\label{bh}
\ee
where ${\mbox{dim}}\, (g)$ is the dimension of $g$, and $r(g)$ is
the rank of the antisymmetric matrix $M_g$ whose elements are  
\be
(M_g)_{ab,cd}={\sum_{m,n=0}^N}
C_{ab,cd}^{mn}\a_{mn} \qquad a<b, \ c<d, \ m<n.
\label{bi}
\ee

For the simple Lie algebras $so(p,q), \ p+q=N+1$, the number $\tau$ is
equal to the rank of the algebra, $\tau=[\frac{N+1}{2}]$. In this case
the upper bound saturates the inequality, and this can be checked by
computing the rank $r(g)$ of the matrix (\ref{bi}). However, it is not
necessary that $g$ be simple in order to have the equality $\tau =
{\mbox{dim}}\,(g) -r(g)$. For instance,   the
Poincar\'e algebra $iso(3,1)$ has two algebraically (quadratic and
fourth-order) independent Casimirs, and in this case the equality also
holds; the same happens for all the inhomogenous pseudo-orthogonal
algebras
${iso}(p,q)$ which appear in the CK family when $\k_1=0$ or $\k_N=0$ and
the remaining ones are different from zero. 

We will show that equality holds in (\ref{bh}) for all CK algebras, so
although the CK set  contains non-simple algebras, all members in each
family have  the same number of algebraically independent Casimirs.
This is the main reason to restrict in the paper from the general set
of graded contractions of $so(N+1)$ to the subfamily of CK algebras. The
property of having exactly $[\frac{N+1}{2}]$ algebraically independent
Casimirs justifies the {\em quasi-simple}  name allocated to them. This
property is no longer true for other contractions of
$so(N+1)$ beyond the CK family, as the extreme case of the abelian
algebra, which has as many primitive Casimirs as generators, clearly
shows.

\subsect{The Gel'fand method}

Instead of solving the set of differential equations (\ref{bf}) in
order to get the elements of $S^I$, Gel'fand considered the
following antisymmetric matrix associated to $so(N+1)$:
\be
T=\left(\begin{array}{cccccc}
0        &  \a_{10} &  \a_{20}   &\dots & \a_{ N-1\,0} & {\a_{N0}} \cr
\a_{01}  &   0      &  \a_{21}   &\dots &  \a_{ N-1\,1} & { \a_{N1}} \cr
\a_{02}  &  \a_{12} &  0         &\dots &  \a_{N-1\,2} & { \a_{N2}} \\ 
\vdots   &  \vdots  &  \vdots    &\ddots& \vdots       & \vdots      \cr
\a_{0\,N-1}&\a_{1\,N-1}&\a_{2\,N-1}&\dots&   0         & \a_{N\,N-1}\cr
\a_{0 N}& \a_{1 N}& \a_{2 N}  &\dots &  \a_{N-1\,N} &   0
\end{array}\right) 
\label{bj}
\ee
where $\a_{ab}=-\a_{ba}$. He obtained the Casimir invariants of 
$so(N+1)$ from the coefficients of the characteristic polynomial of the
matrix $T$:
\be
\det\big(T-\lambda \,{\bf I} \big)=0 ,
\label{bk}
\ee
where ${\bf I}$ is the $(N+1)\times (N+1)$ identity matrix.
Due to the structure of the matrix $T$, these coefficients are
sums of all minors of the same order associated to the main
diagonal of $T$; the last coefficient is of course the determinant of
$T$ \cite{Gelfand}. It turns out that this determinant is equal to
zero when $N$ is even $N=2l$ , and it is a perfect square of an
homogeneous expression of order $l$ in the variables $\a_{ab}$ when
$N$ is odd $N=2l-1$.  The Casimir themselves are obtained through 
the replacement $\a_{ab}\to
\om_{ab}$ and further symmetrization on the variables $\a_{ab}$ 
in these coefficients.

When $N=2l$ is even, Gel'fand obtained $l$ 
such invariants, ${\cal C}_1, \dots, {\cal C}_s,  \dots, {\cal C}_l$ for
$so(N+1)$  which are homogeneous polynomials of order $2s$ in the generators:
\be
{\cal C}_s=\sum_{i_1,i_2,\dots,i_{2s-1},i_{2s} =0}^{N}
\om_{i_1i_2}\om_{i_2i_3}\dots
\om_{i_{2s-1}i_{2s}} \om_{i_{2s}i_1} \qquad s=1,2,\dots,l.
\label{bl}
\ee
When $N$ is odd, $N=2l-1$, besides (\ref{bl}) there is another
invariant coming from the determinant of $T$, which is a
perfect square of an homogeneous expression in the generators of order
$l$ denoted simply as ${\cal C}$:
\be
{\cal C} =
\sum_{i_0,i_1,\dots,i_{N}=0}^{N}\varepsilon_{i_0i_1\dots i_{N}}
\om_{i_0i_1}\om_{i_2i_3}\dots \om_{i_{N-1}i_N},
\label{bm}
\ee
where $\varepsilon_{i_0i_1\dots i_{N}}$ is the completely
antisymmetric unit tensor. In both expressions
the relation $\om_{ab}=-\om_{ba}$ is understood.

\subsect{The case of CK algebras}

We now proceed to implement this scheme for the CK algebras.
  As a first step we write the analogous of $T$
(\ref{bj}) {\em only} for the pseudo-orthogonal $so(p,q)$ algebras
{\em with all $\k_a\ne 0$} (but {\em without} reducing them to $\pm 1$),
and compute the minors separately. Afterwards, we arrange the details, by
introducing some factors depending on the coefficients $\k_a$ in such way
that all contractions $\k_a\to 0$ are always well defined and do not
originate a trivial result for any of the Casimirs found. 

Consider first the CK algebras with all $\k_a \ne 0$. 
Recall that the generators of $so_{\k_1,\dots,\k_N}(N+1)$ have
been taken as $\om_{ab}$ only for $a<b$. If all $\k_a \ne 0$ we
can extend this set and introduce the (linearly dependent)
new generators by defining $\om_{ba}$ with $a<b$, and their
corresponding $\a_{ba}$ as follows:
\be
\mbox{if}\quad a<b \qquad \om_{ba}:= -\frac 1{\k_{ab}}\om_{ab}\qquad
\a_{ba}:= -\frac 1{\k_{ab}}\a_{ab}
\label{ca}
\ee
so that the commutation relations (\ref{ab}) can be written in the
standard form
\be
[\om_{ab},\om_{lm}]=\delta_{am}\om_{lb}-\delta_{bl}\om_{am}+\delta_{bm}\k_{lm}
\om_{al} +\delta_{al}\k_{ab} \om_{bm}.
\label{etiqueta}
\ee
which is the familiar form of the commutation relations for an $so(p,q)$
algebra, with the  non-degenerate metric tensor (\ref{ag}). In this
special case
$\k_a\ne 0$ we associate to $so_{\k_1,\dots,\k_N}(N+1)$ the 
matrix (\ref{bj}) denoted $T_{\k_1,\dots,\k_N}$ which now reads 
\be
T_{\k_1,\dots,\k_N}=\left(\begin{array}{cccccc}
0& -\frac {\a_{01}}{\k_{01}} &-\frac {\a_{02}}{\k_{02}}
&\dots&-\frac {\a_{0N-1}}{\k_{0N-1}} &
-\frac {\a_{0N}}{\k_{0N}} \cr
\a_{01}&0&-\frac {\a_{12}}{\k_{12}}&\dots &-\frac
{\a_{1N-1}}{\k_{1N-1}}  & -\frac {\a_{1N}}{\k_{1N}} \cr
\a_{02}&\a_{12}&0&\dots&-\frac {\a_{2N-1}}{\k_{2N-1}}& 
-\frac {\a_{2N}}{\k_{2N}}
\\ \vdots &\vdots&\vdots&\ddots&\vdots&\vdots  \cr
\a_{0\,N-1}&\a_{1\,N-1}&\a_{2\,N-1}&\dots&0&
 -\frac {\a_{N-1\,N}}{\k_{N-1\,N}} \\
\a_{0 N}&\a_{1 N}&\a_{2 N}&\dots &\a_{N-1\,N}&0
\end{array}\right) .
\label{cb}
\ee
This matrix satisfies the property
\be
T_{\k_1,\dots,\k_N} \, \Ik + \Ik\, {}^t T_{\k_1,\dots,\k_N}=0,
\label{ccn}
\ee
where $\Ik$ is the diagonal matrix (\ref{ag}). Recalling (\ref{cc}), we
say $T_{\k_1,\dots,\k_N}$ is an $\Ik$-antisymmetric matrix.

When {\em all} the constants $\k_a$ are different from zero, it is clear
that the Casimir invariants for
$so_{\k_1,\dots,\k_N}(N+1)$ are the
coefficients of the characteristic polynomial coming from
equation (\ref{bk}) where $T$ is  replaced
by $T_{\k_1,\dots,\k_N}$. In order to get them we have to 
calculate the determinant of a generic diagonal submatrix, which will have
the same structure as (\ref{cb}) but with   a non-consecutive subset of
 indices, say 
$T_{\k_{i_1i_2},\dots,\k_{i_{K-1}i_K}}$. Due to property
(\ref{ccn}) it is easy to show that for an odd $K$ the
determinant is always  zero. Therefore, only determinants for even
$K=2s$ might be different from zero. For future convenience, we will denote
any arrangement of $2s$ indices taken from $012\dots N$  in
increasing order as $a_1\<a_2\<\dots\<a_s\<b_1\<b_2\<\dots\<b_s$. 

We now define some symbols of $2s$ indices $\ww_{a_1a_2\dots
a_sb_1b_2\dots b_s}$, for $s=1,2,\dots, l$ as:
\be
\ww^2_{a_1a_2\dots a_sb_1b_2\dots b_s}
:=\k_{a_1b_s}\k_{a_2b_{s-1}}\dots\k_{a_sb_1}
\det\left[
T_{\k_{a_1a_2},\dots,\k_{a_sb_1},\dots,\k_{b_{s-1}b_s}}
\right].
\label{cd}
\ee
This definition is justified since the r.h.s.\ of (\ref{cd}) is a
perfect square. The set of coefficients $\k_{ab}$ multiplying the
determinant assures that the final expressions
we are going to obtain are non-trivial even after the limits $\k_a\to 0$.
Inserting these $\k$ factors turns out to be equivalent to the
usual rescaling made in the  contraction of Casimir invariants by means
of an In\"on\"u--Wigner contraction.

The $2s$-index $\ww$-symbol  is given in terms of the
$(2s-2)$-index $\ww$-symbol through  
\bea
&& \ww_{a_1a_2\dots a_sb_1b_2\dots b_s} =
  {\sum_{\mu=1}^{s}} (-1)^{\mu+1} \a_{a_\mu b_s}
  \ww_{a_1a_2\dots\widehat{a_\mu}\dots
a_sb_1b_2\dots\widehat{b_s}} \cr 
&& \qquad\qquad\qquad\qquad
  + \ {\sum_{\nu=1}^{s-1}} (-1)^{s+\nu+1} \k_{a_sb_\nu} \a_{b_\nu b_s}
     \ww_{a_1a_2\dots
a_sb_1b_2\dots\widehat{b_\nu}\dots\widehat{b_s}} 
 \label{da}
\eea
where the  $\ww$-symbols  in the r.h.s.\ of the equation
have $2s-2$ indices, those obtained by {\it removing} the two
indices marked with a caret $a_\mu,b_s$ or $b_\nu,b_s$ from the set of
$2s$ indices ${a_1a_2\dots a_sb_1b_2\dots b_s}$.

The $\ww$-symbols  give rise to the elements of $S^I$ (\ref{bf})
and the canonical mapping $\phi$ (\ref{bg}) transforms them
into invariants of the enveloping CK algebra (\ref{be}). The
symmetrization implied by action of  $\phi$ on $\ww$ reduces to a simple
substitution  $\a_{ab}\to \om_{ab}$ since all generators appearing
in the products of the $\ww$-symbols  commute. Once the substitution of
the variables $\a_{ab}$ by the generators
$\om_{ab}$ has been performed, we will denote $W := \phi(\ww)$. Now 
$W_{ab},\  W_{a_1a_2b_1b_2},\ 
\dots,$ are elements in the universal enveloping algebra of the CK
Lie algebra. Their structure can be most clearly presented in a recursive way.
For $a<b$, let:
\be
W_{ab} := \om_{ab},
\ee
then  $W$-symbols  with four indices $a_1 < a_2 <  b_1 < b_2$ are
given in terms of those with two by:
\be 
  W_{a_1a_2b_1b_2} =
  \om_{a_1 b_2} W_{a_2b_1} 
 - \om_{a_2 b_2} W_{a_1b_1}  
 + \k_{a_2b_1} \om_{b_1 b_2} W_{a_1a_2} 
 \label{dan}
\ee 
and further  $W$-symbols  with six, eight, \dots, $2s$ indices, 
$W_{a_1a_2\dots a_sb_1b_2\dots b_s}$  (with $a_1 < a_2 < \dots <
a_s < b_1 < b_2 < \dots <b_s$), are given in terms of those with
two less indices by means of the relations: 
\bea
&& W_{a_1a_2\dots a_sb_1b_2\dots b_s} =
  {\sum_{\mu=1}^{s}} (-1)^{\mu+1} \om_{a_\mu b_s}
  W_{a_1a_2\dots\widehat{a_\mu}\dots
a_sb_1b_2\dots\widehat{b_s}} \cr 
&& \qquad\qquad\qquad\qquad
  + \ {\sum_{\nu=1}^{s-1}} (-1)^{s+\nu+1} \k_{a_sb_\nu} \om_{b_\nu
b_s}
     W_{a_1a_2\dots
a_sb_1b_2\dots\widehat{b_\nu}\dots\widehat{b_s}} 
 \label{daa}
\eea
until we end up with $W$-symbols with $2l$ indices.

Through the use of these $W$'s we can produce
expressions for the Casimir invariants of the CK
algebras ${so}_{\k_1,\dots,\k_N}(N+1)$. The key to this adaptation
is to profit from the presence of the constants $\k_a$. 
When contraction is dealt with  through an
In\"on\"u--Wigner type contraction, a suitable rescaling of the
non-contracted Casimir by some power of the contraction parameter  is
required to give a non-trivial well defined Casimir for the contracted
algebra after the contraction limit. This is made unnecessary in our
approach, which has this rescaling automatically built-in. 

We now give the expressions, in terms of these $W$'s, for
the $[\frac{N+1}{2}]$ Casimir operators in the general CK Lie
algebra ${so}_{\k_1,\dots,\k_N}(N+1)$:

\noindent
 {\it Theorem  2.} The $l=[\frac{N+1}{2}]$ independent polynomial
Casimir operators of the CK Lie algebra
${so}_{\k_1,\dots,\k_N}(N+1)$ can be written as:

\noindent
$\bullet$ $[\frac{N}{2}]$ invariants ${\cal C}_s$, $s=1, \dots, [\frac{N}{2}]$
      of order $2s$. We give the first, the second, and then the
general expression:
\be
{\cal C}_1  =  {\sum_{a_1,b_1=0 \atop 
                  a_1<b_1}^{N}} 
       \k_{0a_1}
       \k_{b_1 N}
       W_{a_1b_1}^2 
\label{dbi}
\ee
\be
{\cal C}_2  = {\sum_{a_1,a_2,b_1,b_2=0 \atop 
                  a_1<a_2<b_1<b_2}^{N}}
       \k_{0a_1}\k_{1a_2}
       \k_{b_1 (N-1)}\k_{b_2 N} 
       W_{a_1a_2b_1b_2}^2 
\label{dbii}
\ee
\be
{\cal C}_s  = \!\!\!\!\!\!\!\!\!\!\!\!\! 
             {\sum_{a_1,a_2,\dots ,a_s,b_1,b_2,\dots, b_s =0 \atop 
                  a_1<a_2<\dots <a_s<b_1<b_2<\dots<b_s }^{N}}
             \!\!\!\!\!\!\!\!\!\!\!\!\!\! 
       \k_{0a_1}\k_{1a_2}  \!\dots\!  \k_{(s-1) a_s}
       \k_{b_1 (N\!-s+\!1)}\k_{b_2 (N\!-s+\!2)}  \!\dots\!  \k_{b_s N} 
       W_{a_1a_2\dots a_sb_1b_2\dots b_s }^2 
\label{db}
\ee

\noindent
$\bullet$ When $N+1$ is even, there is an extra Casimir ${\cal C}$ of
      order $l=\frac{N+1}{2}$: 
\be
{\cal C} = W_{012 \dots N}.
\ee

In these expressions any $\k_{aa}$ should be understood as
$\k_{aa}:=1$. It is easy to see that even in the most contracted
CK algebra, the flag space algebra, ${so}_{0,\dots,0}(N+1)$,
these Casimirs are not trivial. In fact, the term in  
(\ref{db}) with the $W$-symbol   whose first group of $s$ indices
are consecutive and start from $0$, and whose last group of $s$
indices are also consecutive and end with $N$,
$W_{012\dots (s-1) \, (N-s+1)
\dots (N-2) (N-1) N}^2$, is the only term whose $\k$ factor is equal
to 1,  and therefore the only one which survives in the Casimir
${\cal C}_s$ for the
${so}_{0,\dots,0}(N+1)$ algebra.

Therefore, theorem  2 provides a set of $[\frac{N+1}{2}]$
non-trivial independent Casimirs for any Lie algebra in the CK family.
The question now is whether there exists any other Casimir  which cannot be
obtained by  this contraction process. To answer this we should
analyze the upper bound $\tau$ (\ref{bh}).

\noindent
{\it Proposition 3.} The rank of the matrix defined by
(\ref{bi}) for $so_{\k_1,\dots,\k_N}(N+1)$ is
$N^2/2$ for even $N$ and $(N^2-1)/2$ for odd $N$, no matter of the
specific values of the coefficients $\k_a$. 

\noindent
{\em Proof}. The above result is a consequence of the existence of the
structure constants $C_{ab,bc}^{mn}=-\delta_{ma}\delta_{nc}$ (\ref{bd}) 
which are $\k$-independent, and are non-zero for all algebras in the CK 
family. Let us start with the case of an even
$N=2l$ and the $\frac 12 N(N+1)\times \frac 12 N(N+1)$ matrix $M_g$
(\ref{bi}). We consider the minor obtained by eliminating the rows and
columns associated to the following $l=N/2$ variables $\a_{ab}$:
\be
\a_{0N} \quad
\a_{1\,N-1} \quad
\a_{2\,N-2}\quad \dots\quad\a_{l-2\,l+2} \quad\a_{l-1\,l+1},
\label{pro}
\ee
this is, for $\a_{0N}$  we discard  the row and column with elements
$(M_g)_{0N,kl}$ and $(M_g)_{kl,0N}$  ($\forall\, kl$), and so on.
The dimension of this submatrix is $N^2/2$. It can be checked
that  in each row and in each column there is always a {\em
single} $\a$ of the sequence (\ref{pro}) without any factor $\k$.
Then by permuting rows and columns we can  arrange the minor in order to
get all those $\a$'s in the main diagonal, so its determinant is 
(up to a sign), a monomial:
\be
\a_{0N}^{2(N-1)}
\a_{1\,N-1}^{2(N-3)} 
\a_{2\,N-2}^{2(N-5)}  \cdots\a_{l-2\,l+2}^{2\cdot 3}
\a_{l-1\,l+1}^{2\cdot 1}.
\ee
Since this term is $\k$-independent, this minor is non-zero for any CK
algebra. A similar procedure is applied for an odd $N=2l-1$. Now
we  take out the rows and columns linked to the $l=(N+1)/2$ variables:
\be
\a_{0N} \quad
\a_{1\,N-1} \quad
\a_{2\,N-2}\quad \dots\quad\a_{l-2\,l+1} \quad\a_{l-1\,l},
\label{prodos}
\ee
obtaining in this way a minor of dimension $(N^2-1)/2$. By ordering
rows and columns, we get all variables appearing in the sequence
(\ref{prodos}) placed in the main diagonal; the determinant is (again up to
a sign) the monomial:
\be
\a_{0N}^{2(N-1)}
\a_{1\,N-1}^{2(N-3)} 
\a_{2\,N-2}^{2(N-5)}  \cdots\a_{l-3\,l+2}^{2\cdot 4}
\a_{l-2\,l+1}^{2\cdot 2},
\ee
and the result follows.

Hence, as ${\mbox{dim}}\, (g)=\frac 12 N(N+1)$,  the upper bound for
the number of algebraically independent Casimirs in any CK algebra
turns out to be  $[\frac{N+1}{2}]$ which coincides with the upper bound
for the simple algebras where all $\k_a\ne 0$. Since 
we have just obtained that number of Casimirs,   we conclude that 
equality in the formula (\ref{bh})   holds for {\em all} CK algebras  
and theorem 2 gives {\em all} the Casimirs for the CK family. 
 We remark that the   flag algebra $so_{0,\dots,0}(N+1)$ is on
the borderline for this behaviour: if contractions are carried out
beyond this algebra, the rank of the matrix in proposition   3  might
not be given by the same values. This can be easily seen: for the
abelian algebra in $\frac 12 (N+1)N$ dimensions, which can of course
be reached by contracting $so(N+1)$, all generators are central
elements.

If we define now the {\it rank of a Lie algebra} in the CK
family as the number of algebraically independent Casimir
invariants associated to it, then   theorem   2  shows  that all
the CK algebras $SO_{\k_1,\dots,\k_N}(N+1)$ have the same  rank: $N/2$
if $N$ is even and  $(N+1)/2$ if $N$ is odd.

We recall that the first Casimir (\ref{dbi}) for $s=1$ is the
quadratic invariant  related to the Killing--Cartan
form in the case of a simple algebra. If a  ``Killing--Cartan"
form is defined for all CK algebras as usual:
\be
\beta_{ab,cd}\equiv \beta(\om_{ab},\om_{cd})=
\mbox{Trace}\,(\ad \om_{ab}\cdot\ad \om_{cd}) 
={ \sum_{m,n,p,q=0}^N} C_{ ab,mn }^{ pq } 
C_{ cd,pq }^{ mn } 
\label{cj}
\ee
we find, by using the structure constants (\ref{bd}), that in the
basis $\om_{ab}$ this  ``Killing-Cartan" form of the CK algebra
${so}_{\k_1,\dots,\k_N}(N+1)$ is diagonal, and its non-zero
components are
\be
\beta_{ab,ab}=-2(N-1)\,\k_{ab} \qquad a,b=0,\dots,N\qquad a<b.
\label{ck}
\ee
When all the $\k_a\ne 0$ the  Killing--Cartan form is regular (the
algebra is simple or semisimple in the exceptional case $D_2$), so we can
write: 
\be
{\cal C}_1=
\sum_{a,b=0}^N  {\k_{0a}}{\k_{bN}}\om^2_{ab}=\sum_{a,b=0 }^N
\frac{\k_{0N}}{\k_{ab}}\om^2_{ab} =  - 2(N-1) \k_{0N}
\sum_{a,b; c,d=0}^N {\beta^{ab,cd}}\om_{ab}\om_{cd} .
\label{ci}
\ee
This is the known relation giving the quadratic Casimir as the dual
of the Killing--Cartan form, which holds for the case of non-zero
$\k_a$; otherwise the Killing--Cartan form is degenerate, and the last
term in (\ref{ci}) is indeterminate. However, the structure of this
equation shows that in the limit of some
$\k_a\to 0$, while the Killing--Cartan form (\ref{ck}) degenerates, the
Casimir
${\cal C}_1$ remains well defined, because the impossibility of
inverting the matrix $\beta_{ab,cd}$ conspires with the factor $\k_{0N}$ to
produce a well defined limit for ${\cal C}_1$. Similarly, higher order
Casimir invariants are dual to the polarized form of a symmetric
multilinear form.   

The algebraic structure behind the $W$'s which
allows simple expressions for the higher order Casimirs should be
worth studying. For instance, let us consider the commutation
relations among a generator
$\om_{ab}$ and a symbol
$W_{a_1a_2\dots a_{s}b_1b_2\dots b_{s}}$. There are two possibilities:
 
\noindent
(1) If both indices   $a$ and $b$, or none, 
appear in the sequence
$\{a_1a_2\dots a_{s}b_1b_2\dots b_{s}\}$, then the Lie bracket
$[\om_{ab},W_{a_1a_2\dots a_{s}b_1b_2\dots b_{s}}]$ is zero.
 
\noindent
(2) If only {\em one} index $a$ or $b$ belongs to
$\{a_1a_2\dots a_{s}b_1b_2\dots b_{s}\}$, then we have:
\be
[\om_{ab},W_{a_1a_2\dots a_{s}b_1b_2\dots b_{s}}]=(-1)^{p+1}\,
\sqrt{
\k_{ab}\,\frac{\k_{a_1b_{s}}\k_{a_2b_{s-1}}\dots\k_{a_sb_{1}} }{
\k_{a'_1b'_{s}}\k_{a'_2 b'_{s-1}}
\dots\k_{a'_sb'_{1}}}} \,W_{a'_1a'_2\dots
a'_{s}b'_1b'_2\dots b'_{s}}  
\label{cm}
\ee
where the new set of indices $\{a'_1a'_2\dots a'_{s}b'_1b'_2\dots b'_{s}\}$ is
obtained by first writing the sequence 
$\{a,b; a_1a_2\dots a_{s}b_1b_2\dots b_{s}\}$ in increasing order, and
then dropping the common
index. In the factor $(-1)^{p+1}$, 
$p$ means the minimum number of traspositions needed to bring the sequence 
$\{ab;a_1a_2\dots a_{s}b_1b_2\dots b_{s}\}$ into  
increasing order; for instance, in the
sequence $\{13;0234\}\equiv\{0124\}$ $p=3$, while in $\{15;1234\}\equiv\{2345\}$
$p=4$. Notice that all $\k_{a'b'}$ in the denominator cancel, and leave under
the square root a perfect square product of $\k$'s.  

Repeated use of this procedure would give the commutation relations
between two $W$-symbols. We do not write here the general
expressions, but some examples are given in the next section.


\sect{Examples}

\subsect{Results for N=2,3,4,5}

In the sequel we elaborate upon the results of the above section
by  writing explicitly  the Casimir invariants of 
$so_{\k_1,\dots,\k_N}(N+1)$ up to $N=5$. These expressions should
be compared with the ones obtained in any approach giving the 
invariants directly in terms of the generators $\om_{ab}$, without
using the $W$-symbols, which are cumbersome as soon as $N$
grows. 

\medskip

\noindent
${\bf{ N=2}}$. There is only one invariant:
\be
{\cal C}_1=\k_{2}\om_{01}^2+\om_{02}^2+\k_{1}\om_{12}^2.
\label{daii}
\ee

\medskip

\noindent
${\bf{ N=3}}$. There are two invariants and  the first relevant
$W$-symbol (\ref{dan})  appears: 
\be
\begin{array}{ccc}
{\cal C}_1=
\k_2\k_3\om_{01}^2 & +\k_{3}\om_{02}^2    & +\om_{03}^2     \cr 
                    & +\k_1\k_3 \om_{12}^2 & +\k_1\om_{13}^2 \\
                    &                        & +\k_1\k_2\om_{23}^2  
\label{dbiii}
\end{array}
\ee
\bea 
{\cal C}\equiv W_{0123}\!\!&=&\!\!\k_{12}\om_{23}W_{01} -\om_{13} W_{02} +
\om_{03}W_{12}\cr 
&=&\!\!\k_{2}\om_{23}\om_{01} -\om_{13} \om_{02} +
\om_{03}\om_{12}.\label{dc}
\eea

\medskip
 
\noindent
${\bf{ N=4}}$. From a physical point of view this is a rather interesting
case since the CK family $so_{\k_1,\dots,\k_4}(5)$ contains the (3+1)
kinematical algebras. There are two invariants:
\be
\begin{array}{cccc}
{\cal C}_1=\k_2\k_3\k_4 \om_{01}^2 & +\k_3\k_4 \om_{02}^2 &
+\k_4\om_{03}^2 & +\om_{04}^2 \cr
  & +\k_1\k_3\k_4\om_{12}^2 & +\k_1\k_4\om_{13}^2 & +\k_1\om_{14}^2\\ 
  &  & +\k_1\k_2\k_4 \om_{23}^2 & +\k_1\k_2  \om_{24}^2  \cr
  &  &  & + \k_1\k_2\k_3 \om_{34}^2 
\label{dd}
\end{array}
\ee
\be
{\cal C}_2=\k_{24}W_{0123}^2+\k_{23}W_{0124}^2+W_{0134}^2+\k_{12}W^2_{0234}
+\k_{02}W^2_{1234} 
\label{de}
\ee
where from (\ref{dan}) we have 
\be
\begin{array}{l}
 W_{0123}= \k_{12}\om_{23}\om_{01} -\om_{13} \om_{02} +
\om_{03}\om_{12} \cr
 W_{0124}= \k_{12}\om_{24}\om_{01} -\om_{14} \om_{02} +
\om_{04}\om_{12} \cr
W_{0134}= \k_{13}\om_{34}\om_{01} -\om_{14} \om_{03} +
\om_{04}\om_{13} \cr
W_{0234}= \k_{23}\om_{34}\om_{02} -\om_{24} \om_{03} +
\om_{04}\om_{23} \cr
W_{1234}= \k_{23}\om_{34}\om_{12} -\om_{24} \om_{13} +
\om_{14}\om_{23} .
\end{array}
\label{df}
\ee

As an application of  (\ref{cm}) we write the
non-zero commutation relations among the ten generators of
$so_{\k_1,\k_2,\k_3,\k_4 }(5)$ and the five $W$-symbols (\ref{df}):
$$
\begin{array}{lll}
 [\om_{04},W_{0123}]=-\k_{02}W_{1234}&\  
[\om_{03},W_{0124}]=\k_{02}W_{1234}&\   
[\om_{02},W_{0134}]=-\k_{02}W_{1234}\cr
 [\om_{14},W_{0123}]=\k_{12}W_{0234}&\   
[\om_{13},W_{0124}]=-\k_{12}W_{0234}&\   
[\om_{12},W_{0134}]=\k_{12}W_{0234}\cr
 [\om_{24},W_{0123}]=-W_{0134}&\     
[\om_{23},W_{0124}]= W_{0134}&\   
[\om_{23},W_{0134}]=- \k_{23}W_{0124}\cr
 [\om_{34},W_{0123}]= W_{0124}&\    
[\om_{34},W_{0124}]=-\k_{34}W_{0123}&\      
[\om_{24},W_{0134}]=\k_{24}W_{0123} 
\end{array}
\nonumber
$$
\be
\begin{array}{ll}
 [\om_{01},W_{0234}]=\k_{01}W_{1234}&\quad  
[\om_{01},W_{1234}]= - W_{0234}  \cr
 [\om_{12},W_{0234}]=-W_{0134}& \quad
[\om_{02},W_{1234}]=  W_{0134} \cr
 [\om_{13},W_{0234}]=\k_{23}W_{0124}&\quad 
[\om_{03},W_{1234}]= - \k_{23} W_{0124} \cr
 [\om_{14},W_{0234}]=-\k_{24}W_{0123}&\quad 
[\om_{04},W_{1234}]=  \k_{24} W_{0123} 
\end{array}
\label{dh}
\ee

As these expressions show,  the $W$-symbols
can be thought of as a kind of ``Pauli--Lubanski" components for
Lie algebras in the CK family. To see this, notice that when the
constants $(\k_1,\k_2,\k_3,\k_4)\equiv
(0,-1,+1,+1)$, the CK Lie algebra $so_{0,-1,+1,+1}(5)$ is isomorphic to
the Poincar\'e algebra $iso(3,1)$. In this case there are {\em five}
$W$-symbols; due to the vanishing of the constant $\k_1$, $W_{1234}$ is missing 
in the expression of the fourth-order Casimir (\ref{de}), and 
the  remaining {\em four} are the components
of the standard Pauli--Lubanski operator (see (\ref{dp}) below). 

From (\ref{dh}) is
straightforward to find the Lie brackets of the $W$-symbols among themselves:
\be
\begin{array}{l}
 [W_{0123},W_{0124}] =\k_{12}\om_{01}W_{0134} +\k_{12}\om_{02}W_{0234}
+\k_{02}\om_{12} W_{1234} \cr
 [W_{0123},W_{0134}] =-\k_{13}\om_{01}W_{0124} +\k_{12}\om_{03}W_{0234}
+\k_{02}\om_{13} W_{1234} \cr
 [W_{0123},W_{0234}] =-\k_{23}\om_{02}W_{0124} - \om_{03}W_{0134}
+\k_{02}\om_{23} W_{1234}\cr
 [W_{0123},W_{1234}] =-\k_{23}\om_{12}W_{0124} - \om_{13}W_{0134}
-\k_{12}\om_{23} W_{0234}\cr
 [W_{0124},W_{0134}] =\k_{14}\om_{01}W_{0123} + \k_{02}\om_{14}W_{1234}
+\k_{12}\om_{04} W_{0234}\cr
 [W_{0124},W_{0234}] =\k_{24}\om_{02}W_{0123} - \om_{04}W_{0134}
+\k_{02}\om_{24} W_{1234}\cr
 [W_{0124},W_{1234}] =\k_{24}\om_{12}W_{0123} - \om_{14}W_{0134}
-\k_{12}\om_{24} W_{0234}\cr
 [W_{0134},W_{0234}] =\k_{24}\om_{03}W_{0123} + \k_{23}\om_{04}W_{0124}
+\k_{03}\om_{34} W_{1234}\cr
 [W_{0134},W_{1234}] =\k_{24}\om_{13}W_{0123} +\k_{23} \om_{14}W_{0124}
-\k_{13}\om_{34} W_{0234}\cr
 [W_{0234},W_{1234}] =\k_{24}\om_{23}W_{0123} + \k_{23}\om_{24}W_{0124}
+\k_{23}\om_{34} W_{0134}
\end{array}
\label{di}
\ee

\medskip
 
\noindent
${\bf{ N=5}}$. The three
invariants are given by:
\be
\begin{array}{ccccc}
{\cal C}_1=\k_{15} \om_{01}^2 & +\k_{25}
\om_{02}^2 & +\k_{35}\om_{03}^2 & +\k_{45}\om_{04}^2 & +\om_{05}^2\cr
 & + 
\k_{01}\k_{25} \om_{12}^2 & +\k_{01}\k_{35} \om_{13}^2 & +
\k_{01}\k_{45} \om_{14}^2 & +\k_{01}\om_{15}^2 \cr
 &  &  +\k_{02}\k_{35} \om_{23}^2 & +
\k_{02}\k_{45} \om_{24}^2 & +\k_{02}\om_{25}^2\\
 &  &  &  +
\k_{03}\k_{45}
\om_{34}^2 & +\k_{03}\om_{35}^2  \cr
 &  &  &  & +\k_{04}\om_{45}^2 
\label{dj}
\end{array}
\ee
\be
\begin{array}{l}
{\cal C}_2=\k_{35}\k_{24}W^2_{0123}+\k_{25}W^2_{0124}
+\k_{24}W^2_{0125}+\k_{35}W^2_{0134}+\k_{34}W^2_{0135}\cr
 \quad +
W^2_{0145}+\k_{12}\k_{35}W^2_{0234}+\k_{12}\k_{34}W^2_{0235}+
\k_{12} W^2_{0245}+\k_{13}W^2_{0345}\cr
 \quad +
\k_{02}\k_{35}W^2_{1234}+\k_{02}\k_{34}W^2_{1235}+\k_{02}W^2_{1245}+
\k_{03}W^2_{1345}+\k_{02}\k_{13}W^2_{2345} 
\label{dk}
\end{array}
\ee
\be
{\cal C}\equiv
W_{012345}=\k_{24}\om_{45}W_{0123}-\k_{23}\om_{35}W_{0124}
+\om_{25}W_{0134}-\om_{15}W_{0234}
+\om_{05}W_{1234}.
\label{dl}
\ee
 
In the above expressions it can be noticed  how all the contractions
$\k_a\to 0$ are always well defined and lead to non-trivial results.
For the most contracted algebra 
$so_{0,0,0,0,0}(6)$, for instance, we get:
\be
{\cal C}_1=\om_{05}^2,  \qquad 
{\cal C}_2=W^2_{0145},  \qquad 
{\cal C}=W_{012345}.
\ee
And for  $N$ arbitrary, the Casimirs   for this most contracted CK
algebra are: 
\be
{\cal C}_1=\om_{0N}^2 \equiv W_{0N}^2, \qquad 
{\cal C}_2=W^2_{01(N-1)N},  \qquad 
{\cal C}_3=W^2_{012(N-2)(N-1)N}, \  \dots
\label{Wflag}
\ee
In fact, even if the appearance of the factors $\k_{ab}$ 
seems rather haphazard when written in specific cases,
like in (\ref{dk}), they are indeed easily reconstructed from
scratch, without reference to the general expressions (\ref{db}).
If here $[X]$ denotes the dimension of $X$, dimensional homogeneity of
commutation relations requires that
$[\om_{ab}]=[\k_{ab}]^{1/2}$, so $[W_{ab}^2]=[\k_{ab}]$,
$[W_{a_1a_2b_1b_2}^2]=[\k_{a_1b_2}][\k_{a_2b_1}]$, etc. By simply
recalling the $W$ term (\ref{Wflag}) entering into any Casimir {\em
without\/} any $\k_{ab}$ factor, then the coefficient of any other $W^2$ is
unambiguously derived by simply requiring dimensional homogeneity for
all $W^2$ terms in the Casimirs and recalling that all terms enter with the
same global sign there. Signs coming from signatures which appear in the
standard known cases (in the Minkowski square of the Pauli-Lubanski vector, for
instance), are hidden inside the $\k_{ab}$ themselves.

\subsect{Casimirs for 3+1 kinematical algebras}

After this algebraic description of the structure of the invariants of the
CK algebras we focus on the most important kinematical algebras
included in the $so_{\k_1,\k_2,\k_3,\k_4}(5)$ family. 
Let $H$, $P_i$, $K_i$ and $J_i$ $(i=1,2,3)$ the usual generators of time
translation, space translations, boosts and spatial rotations,
respectively. Under the following identification with the ``abstract"
generators $\om_{ab}$:
\bea
&&H=\om_{01} \qquad  P_i=\om_{0\, i+1} \qquad
 K_i=\om_{1\, i+1} \qquad i=1,2,3 \cr
&&J_1=\om_{34} \qquad J_2=-\om_{24}  \qquad J_3=\om_{23}  
\label{dm}
\eea
we can interpretate the six CK algebras $so_{\k_1,\k_2,+,+}(5)$ with
$\k_2\le 0$ as the Lie algebras  of the groups of motions of different
(3+1) spacetime models \cite{BLL}.  Note that we have fixed the two
coefficients $\k_3$ and $\k_4$ to
$+1$ (this is a consequence of the space isotropy). The two remaining ones
have a definite physical interpretation:
$\k_1$ is the constant curvature of the spacetime which
appears here as the homogeneous space given in (\ref{ah}), which is the
quotient ${\cal
X}_0=SO_{\k_1,\k_2,+,+}(5)/SO_{\k_2,+,+}(4)$, where 
$SO_{\k_2,+,+}(4)$ is the subgroup generated by the subalgebra
$h_0=\langle K_i,J_i\rangle$: this is the subalgebra of isotopy of a point
in spacetime. Similarly,
$\k_2$ is the curvature of the space of time-like lines in spacetime, 
 ${\cal
X}_{01}=SO_{\k_1,\k_2,+,+}(5)/(SO_{\k_1}(2)\otimes SO_{+,+}(3))$, where
now $SO_{\k_1}(2)\otimes SO_{+,+}(3)$ is the isotopy subgroup of a
time-like line, generated by
$h_{01}=\langle H,J_i\rangle$. This  curvature is linked to the
fundamental constant
$c$ of relativistic theories as
$\k_2=\frac{-1}{c^2}$. To make the comparison easier for these
cases, we shall write the two constants involved in the
``kinematical" subfamily of CK algebras as:
$\kk\equiv\k_1$, $\frac{-1}{c^2}\equiv\k_2$.

In this notation the commutation rules (\ref{ab}) of
$so_{\kk,-1/c^2,+,+}(5)$ read now:
\bea
&&\back\back [J_i,J_j]=\varepsilon_{ijk}J_k\qquad
[J_i,P_j]=\varepsilon_{ijk}P_k\qquad
[J_i,K_j]=\varepsilon_{ijk}K_k\cr
&&\back\back [P_i,P_j]=-\frac {\kk}{c^2}\varepsilon_{ijk}J_k
\quad [K_i,K_j]=-\frac {1}{c^2}\varepsilon_{ijk}J_k
\quad [P_i,K_j]=-\frac {1}{c^2}\delta_{ij}H\\
 &&\back\back [H,P_i]=\kk K_i\qquad [H,K_i]=-P_i\qquad
[H,J_i]=0\qquad i,j,k=1,2,3.
\nonumber
\eea

The limit
$\kk\to 0$ (space-time contraction)  gives rise to the {\em flat}
universes (Min\-kows\-ki and Galilei) coming from the curved ones
(DeSitter and Newton-Hooke); in terms of the ``universe radius"
$R:=\frac{1}{\sqrt\kk}$ or
$R:=\frac{1}{\sqrt{-\kk}}$, this is usually made as $R\to\infty$.
The limit $c\to \infty$ (speed-space
contraction) leads to ``absolute-time" spacetimes (Newton-Hooke and
Galilei) coming from ``relative-time" ones (DeSitter and
Minkowski). 

In this context, the Casimir invariants  (\ref{dd}) and (\ref{de}) adopt
the form
 \be 
 {\cal
C}_1=P_1^2+P_2^2+P_3^2-\frac{1}{c^2}H^2+\kk(K_1^2+K_2^2+K_3^2)
-\frac{\kk}{c^2}(J_1^2+J_2^2+J_3^2) 
\label{dn}
\ee
\be
{\cal C}_2= W_{0123}^2+ W_{0124}^2+W_{0134}^2-\frac{1}{c^2}W^2_{0234}
-\frac{\kk}{c^2}W^2_{1234} 
\label{do}
\ee
where
\bea
&&W_{0123}=-\frac{1}{c^2}HJ_3 -P_1 K_2+ P_2K_1 \cr
&& W_{0124}=\frac{1}{c^2}HJ_2 -P_1 K_3+ P_3K_1 \cr
&& W_{0134}=-\frac{1}{c^2}HJ_1 -P_2 K_3+ P_3K_2\label{dp}\\
&& W_{0234}= P_1J_1 +P_2 J_2+ P_3J_3 \cr
&& W_{1234}= K_1J_1 +K_2 J_2+ K_3J_3.
\nonumber
\eea

We display in table I these six kinematical algebras together their
invariants according to the values of $(\kk,\frac{-1}{c^2},+,+)$.
The limit transitions among them can be clearly appreciated; note that some
$W$-symbols (\ref{dp}) are ``internally contracted" in the case of  $c=\infty$.

\bigskip

{
\noindent
{\bf {Table I}}.  The Casimir invariants of $so_{\kk,-1/c^2,+,+}(5)$. 
\smallskip \footnotesize

\noindent
\begin{tabular}{|l|l|l|}
\hline
Oscillating Newton--Hooke&Galilei&Expanding Newton--Hooke\\
$(+,0,+,+)$\hfill$\kk=1, c=\infty$&
$(0,0,+,+)$\hfill$\kk=0, c=\infty$&
$(-,0,+,+)$\hfill$\kk=-1, c=\infty$\\
$t_6(so(3)\oplus so(2))$&$iiso(3)$&$t_6(so(3)\oplus so(1,1))$\\
\hline
${\cal C}_1=P_1^2+P_2^2+P_3^2$&${\cal C}_1=P_1^2+P_2^2+P_3^2$&
${\cal C}_1=P_1^2+P_2^2+P_3^2$\\
$\qquad +  K_1^2+K_2^2+K_3^2 $& &$\qquad -  K_1^2-K_2^2-K_3^2 $\\
${\cal C}_2= W_{0123}^2+ W_{0124}^2+W_{0134}^2$&
${\cal C}_2= W_{0123}^2+ W_{0124}^2+W_{0134}^2$&
${\cal C}_2= W_{0123}^2+W_{0124}^2+W_{0134}^2$\\
\hline
\hline
Anti-DeSitter\quad $so(3,2)$&Poincar\'e\quad $iso(3,1)$&DeSitter\quad
$so(4,1)$\\
$(+,-,+,+)$\hfill$\kk=1, c=1$&
$(0,-,+,+)$\hfill$\kk=0, c=1$&
$(-,-,+,+)$\hfill$\kk=-1, c=1$\\
\hline
${\cal C}_1=P_1^2+P_2^2+P_3^2-H^2$&${\cal C}_1=P_1^2+P_2^2+P_3^2-H^2$&
${\cal C}_1=P_1^2+P_2^2+P_3^2-H^2$\\
$\qquad +  K_1^2+K_2^2+K_3^2 $& &$\qquad -  K_1^2-K_2^2-K_3^2 $\\
$\qquad - J_1^2-J_2^2-J_3^2 $& &$\qquad +J_1^2+J_2^2+J_3^2 $\\
${\cal C}_2= W_{0123}^2+ W_{0124}^2+W_{0134}^2$&
${\cal C}_2= W_{0123}^2+ W_{0124}^2+W_{0134}^2$&
${\cal C}_2= W_{0123}^2+W_{0124}^2+W_{0134}^2$\\
$\qquad -W^2_{0234} -W^2_{1234}$&
$\qquad -W^2_{0234} $&
$\qquad -W^2_{0234} + W^2_{1234}$\\
\hline
\end{tabular} }
 
\bigskip


\sect{Concluding remarks}

The Casimir invariants play a prominent role in any problem where a
Lie algebra and its enveloping algebra appear. One example of
current active interest is the theory of quantum groups; explicit
deformations of the
$W$-symbols can be found in the deformed commutation relations of  the
quantum  CK algebras  $U_z so_{0,\k_2,\dots,\k_N}(N+1)$ with $\k_1=0$, and
indeed the study of Casimirs in the classical undeformed algebras we have
presented here underlies the expressions for the deformed Casimirs
in \cite{afin}.

A more classical application concerns the expansions processes which can be
seen as the opposite situation of a contraction limit \cite{Gilmore}. 
While the contractions makes to vanish some structure constants of a Lie
algebra  which therefore gets more abelian, the expansions start from a Lie
algebra, with some Lie brackets typically equal to zero, and ends up with
another less abelian algebra, which is usually realized as a Lie
subalgebra of the universal enveloping algebra of the initial Lie
algebra. The more known transitions of this kind are the rank-one
expansions which allows to obtain the simple
$so(p,q)$ algebras starting from inhomogeneous $iso(p,q)$; the name
rank-one refers to the rank of the homogeneous spaces behind these
expansions (the constant curvature flat and curved space with a metric of
signature
$(p,q)$)  \cite{Gilmore,Rosen}. In our framework this fact is equivalent to
``create" a non-zero coefficient $\k_1$ out of the case $\k_1=0$. 
Technically, these expansions only involve the quadratic Casimir
${\cal C}_1$ to get the correct Lie subalgebra in the enveloping algebra
to be expanded (within an irreducible representation). Extension of
these results for high-order expansions would be of
great interest. The rank-two  expansions would go from
$t_r(so(p,q)\oplus so(p',q'))$ (Newton--Hooke algebras) to
$so(p,q)$ algebras, this is, they would introduce curvature into the space
of lines, just in the same way as the rank-one expansions go from
flat to curved spaces of points. Here the two first Casimir
invariants
${\cal C}_1$ and
${\cal C}_2$ should participate, and it is reasonable to guess that the
explicit introduction of the constants $\k_a$ may help in the choice of
the correct expansion procedure which, as far as we know, is yet unknown 
for higher rank spaces.



\bigskip
\bigskip

\noindent
{\Large{{\bf Acknowledgments}}}

\bigskip

 This work has been
partially supported by DGICYT (Project PB94--1115) from the
Ministerio de Educaci\'on y Ciencia de Espa\~na.

\bigskip


\end{document}